\documentclass[10pt]{forma}
\def\insertplot#1{\includegraphics{#1}}

\def\be{\begin{equation}}
\def\ee{\end{equation}}

\def\bea{\begin{eqnarray}}
\def\eea{\end{eqnarray}}

\begin{document}
%%%%%%%%%%%%%%%%%%%%%%%%%%%%%%%%%%%%%%%%%%

\hspace{1.0cm} \parbox{15.0cm}{

\baselineskip = 15pt

\noindent {\bf FACTOR ANALYSIS OF GALACTIC GLOBULAR CLUSTERS ON STRUCTURAL PARAMETERS}

%%%%%%%%%%%%%%%%%%%%%%%%%%%%%%%%%%%%%%%%%%
\bigskip
\bigskip

\noindent {\bf O.Eigenson$^1$, O.Yatsyk$^1$}

%%%%%%%%%%%%%%%%%%%%%%%%%%%%%%%%%%%%%%%%%%
\bigskip

\baselineskip = 9.5pt

\noindent {\small \copyright~2000}

\smallskip

$^1$\noindent {\small {\it  Astronomical Observatory of Ivan Franko national university of 
Lviv, Lviv, Ukraine}} \\

\noindent {\small {\it e-mail:}} {\tt yatsyk@astro.franko.lviv.ua}

%%%%%%%%%%%%%%%%%%%%%%%%%%%%%%%%%%%%%%%%%%
\baselineskip = 9.5pt \medskip

\medskip \hrule \medskip

\noindent Principal component method is used to study galactic globular clusters in 7- 
and 8-axis space of structural parameters. It is shown that the manifold 
properties of clusters with this set of parameters is determined mainly by 
two independent factors. This result may be useful for the theory of formation 
and evolution of clusters.

\medskip \hrule \medskip

}
%%%%%%%%%%%%%%%%%%%%%%%%%%%%%%%%%%%%%%%%%%
\vspace{1.0cm}

\baselineskip = 11.2pt
%%%%%%%%%%%%%%%%%%%%%%%%%%%%%%%%%%%%%%%%%%

\bigskip

In authors paper \cite{EY86} the factor analysis (principal 
component method) of globular clusters on some main physical parameters
was performed. It was shown that the manifold properties of clusters is
determined mainly by two independent factors. The first of them is metallicity
which is connected with the distance to Galaxy center, and the second one
reflects the richness of the cluster. Analogous studies were performed by \cite{BL84}, 
and later by \cite{DM94} and \cite{D95}.

In this paper we return to this problem with some other set of parameters,
different not only from our former set but also from the set used by Djorgovski 
and Meylan. The basedata was the catalogue of 125 clusters of \cite{TKD95}
with the following parameters of clusters: concentration $\it c$,
core-radius $\it r_{c}$, a half-light radius $\it r_{h}$, and other fraction-of-light
radii $\it r_{10}$, $\it r_{20}$, $\it r_{30}$, $\it r_{40}$. In the other case the quantity 
$\it \mu_V{(0)}$, central surface brightness, was also included. The details of 
calculation one may find in our cited paper, or in the paper of \cite{MH86}.

The correlation matrix is given in tab.1.

%#################################################################%
\begin{table}[htbp]

{\footnotesize

\caption{Matrix of the correlation coefficients.}
\label{tab1}
\begin{center}
\begin{tabular}{|l|rrrrrrr|}
\hline
&$\it r_{c}$ &$\it r_{10}$ &$\it r_{20}$ &$\it r_{30}$ &$\it r_{40}$ &$\it r_h$ &$\it c$\\
\hline
$\it r_{c}$ &1.000 &.848 &.773 &.712 &.661 &.619 &-.687\\
$\it r_{10}$ &.848 &1.000 &.989 &.971 &.951 &.930 &-.420\\
$\it r_{20}$ &.773 &.989 &1.000 &.994 &.983 &.969 &-.331\\
$\it r_{30}$ &.712 &.971 &.994 &1.000 &.997 &.989 &-.262\\
$\it r_{40}$ &.661 &.951 &.983 &.997 &1.000 &.997 &-.207\\
$\it r_h$ &.619 &.930 &.969 &.989 &.997 &1.000 &-.163\\
$\it c$ &-.687 &-.420 &-.331 &-.262 &-.207 &-.163 &1.000\\
\hline
\end{tabular}
\end{center}

}

\end{table}
%#################################################################%

\bigskip

Eigenvalues (characteristic roots) are the following: $\lambda_1$ = 5.635,
$\lambda_2$ = 1.180, $\lambda_3$ = 0.170, $\lambda_4$ = 0.012, $\lambda_5$ = 0.002,
$\lambda_6$ = 0.001, $\lambda_7$ = 0.000. We see that these root decrease rapidly 
enough, and one may be limited by the first two roots. The matrix of factor
projections is the following (tab.2):

\bigskip
%#################################################################%
\begin{table}[htbp]

{\footnotesize

\caption{Matrix of the factor projections.}
\label{tab2}
\begin{center}
\begin{tabular}{|l|rrrrrrr|}
\hline
&$\it r_{c}$ &$\it r_{10}$ &$\it r_{20}$ &$\it r_{30}$ &$\it r_{40}$ &$\it r_{h}$ &$\it c$\\
\hline      
$F_1$ &0.825 &0.996 &0.994 &0.982 &0.967 &0.950 &-0.415\\                                        
$F_2$ &-0.472 &-0.023 &0.094 &0.178 &0.241 &0.289 &0.880\\
\hline                                       
\end{tabular}
\end{center}

}

\end{table}
%#################################################################%

This may be illustrated by the Fig.1.

%************************************************************************%
\begin{figure}[htb]
\vspace{5.0cm}
\insertplot{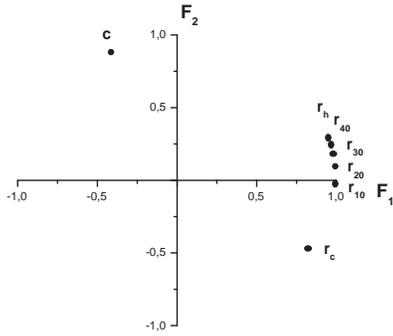}
\caption{Diagram $F_{1} - F_{2}$ for 7 parameters.}
\label{fig1}
\end{figure}
%************************************************************************%

If we include the central surface brightness $\it \mu_V{(0)}$, the picture becomes more
complicated (Fig.2). So this brightness is determined simultaneously by both two factors.

%************************************************************************%
\begin{figure}[htb]                  
\vspace{5.0cm}                       
\insertplot{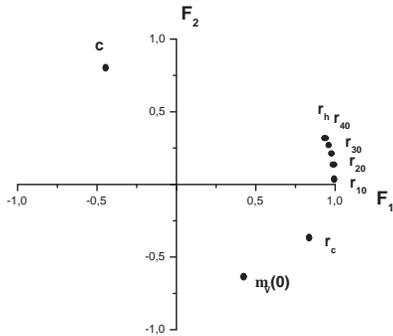}     
\caption{Diagram $F_{1} -F_{2}$ for 8 parameters.}
\label{fig2}
\end{figure}                         
%************************************************************************%

We see that the manifold properties of clusters is determined mainly by two independent
factors. This result may be useful for the theory of formation and evolution of
globular clusters.

%%%%%%%%%%%%%%%%%%%%%%%%%%%%%%%%%%%%%%%
\bigskip
\noindent {\small {\bf REFERENCES}}

%%%%%%%%%%%%%%%%%%%%%%%%%%%%%%%%%%%%%%%%%%

%%%%%%%%%%%%%%%%%%%%%%%%%%%%%%%%%%%%%%%%%%

\vspace{-1.0cm}

\begin{thebibliography}{99}

\small{
\baselineskip=1pt

\bibitem{BL84}
{\it Broche P., Lendes, F.} The manifold of globular clusters 
// Astron. and Astrophys.--1984.--{\bf 139}, N.2.--P.474-476. 

\bibitem{EY86}
{\it Eigenson A.M., Yatsyk O.S.} Factor analysis of globular 
clusters - the principal components method // Soviet Astr.--
1986.--{\bf 30}, N 4.--P. 390-394.

\bibitem{DM94}
{\it Djorgovski S., Meylan G.} The galactic globular cluster 
system // Astron.J.--1994.--{\bf 108}.--P. 1292-1311.

\bibitem{D95}
{\it Djorgovski S.} The fundamental plane correlations for 
globular clusters // Astrophys. J.--1995.--{\bf 438}, N.1.--P.
L29-L32.

\bibitem{MH86}
{\it Murtagh F., Heck A.} An annotated bibliographical catalogue
of multivariate statistical methods and of their astronomical 
applications // Bull.Inf.Centre Donnees Stellaires --1986.--
{\bf 31}.--P.183.

\bibitem{TKD95}
{\it Trager S.C., King I.R., Djorgovski S.} Catalogue of galactic 
globular-cluster surface-brightness profiles // Astron. J.--1995.--
{\bf 109}.--P.218-241.
 
}

\end{thebibliography}
\end{document}